\documentclass[12pt]{article}
\usepackage[colorlinks,linkcolor=blue,citecolor=blue,urlcolor=blue,bookmarks,bookmarksnumbered]{hyperref}
\usepackage{mathpazo,charter}
\usepackage[scaled=0.95]{helvet}
\usepackage{amsmath,XXXE}
\usepackage{graphicx,color}
\usepackage{booktabs,multirow,array}

\def\dt#1{\accentset{\hbox{\normalsize.}}{#1}}
\def\ddt#1{\accentset{\hbox{\normalsize\kern.5pt.\kern-1pt.}}{#1}}
\def\vdt{\partial_\tau}

\definecolor{Green}  {rgb}{0.10,0.70,0.10} 
\definecolor{Orange} {rgb}{1.00,0.50,0.15} 
\definecolor{Red}    {rgb}{0.90,0.00,0.12} 
\definecolor{Purple} {rgb}{0.60,0.22,0.65} 
\definecolor{Turque} {rgb}{0.00,0.65,0.85} 
\definecolor{Blue}   {rgb}{0.00,0.00,1.00} 
\definecolor{Magenta}{rgb}{1.00,0.00,1.00} 
\definecolor{Gold}   {rgb}{1.00,0.75,0.25} 
\definecolor{Seaweed}{rgb}{0.02,0.43,0.16} 
\definecolor{Brown}  {rgb}{0.43,0.26,0.32} 
\definecolor{grey1}  {rgb}{0.20,0.20,0.20} 
\definecolor{grey2}  {rgb}{0.40,0.40,0.40} 
\definecolor{grey3}  {rgb}{0.60,0.60,0.60} 
\definecolor{grey4}  {rgb}{0.80,0.80,0.80} 
\definecolor{grey5}  {rgb}{0.90,0.90,0.90} 
\def\C#1#2{{\ifcase#1\or
             \color{Green}\or \color{Orange}\or \color{Red}\or
              \color{Purple}\or \color{Turque}\or \color{Blue}\or
               \color{Magenta}\or \color{Gold}\or \color{Seaweed}\or
                \color{Brown}\or\color{grey1}\or\color{grey2}\or
                 \color{grey3}\else\color{grey4}\fi#2}}
\def\cb#1#2{\setlength\fboxsep{1pt}\colorbox{#1}{\color{#1}\fbox{\color{black}#2}}}
\def\cB#1{\hbox to0pt{\setlength\fboxsep{0pt}\hss\color{grey3}\fbox{\cb{white}{#1}}\hss}}
\def\bB#1{\hbox to0pt{\setlength\fboxsep{0pt}\hss\color{grey3}\fbox{\cb{black}{\color{white}#1}}\hss}}

 \SfTitles
 \allowdisplaybreaks

\begin{document}

\thispagestyle{empty}
 \noindent
 \today\hfill
  \vspace*{5mm}
 \begin{center}
{\LARGE\sf\bfseries A Q-Continuum of Off-Shell Supermultiplets}\\[1mm]
  \vspace*{5mm}
 \begin{tabular}{p{80mm}cp{80mm}}
 \hfill
{\large\sf\bfseries  Tristan H\"{u}bsch$^{*\dag}$}
                    &and&
{\large\sf\bfseries  Gregory A.~Katona$^{\dag\ddag}$}\\[1mm]
\MC3c{\small\it
  $^*$\,Department of Physics \&\ Astronomy,
  Howard University, Washington, DC 20059} \\[-1mm]
\MC3c{\small\it
  $^\dag$\,Department of Physics, University
  of Central Florida, Orlando, FL 32816}\\[-1mm]
\MC3c{\small\it
  $^\ddag$\,Affine Connections, LLC, College Park, MD 20740}\\[0mm]
 \hfill {\tt  thubsch@howard.edu}&&{\tt grgktn@knights.ucf.edu}
  \end{tabular}\\[1mm]
  \vspace*{5mm}
{\sf\bfseries ABSTRACT}\\*[3mm]
\parbox{148mm}{We explore a continuum of observably and usefully inequivalent, finite-dimensional off-shell representations of worldline $N\,{=}\,4$-extended supersymmetry, differing from one another only in the value of a ``tuning parameter.'' Their dynamics turns out to be nontrivial already when restricting to just bilinear Lagrangians. In particular, we find a 34-parameter family of bilinear Lagrangians that couple two differently ``tuned'' of these supermultiplets to each other and to external magnetic fields, where the explicit dependence on the tuning parameters cannot be removed by any field redefinition and so is observable.
 } 
\end{center}
\vspace{5mm}
\noindent
\parbox[t]{60mm}{PACS: {\tt11.30.Pb}, {\tt12.60.Jv}}\hfill
\parbox[t]{100mm}{\raggedleft\small\baselineskip=12pt\sl
             Discreteness is the refuge of the clumsy.\\[-0pt]
            |~Jorge Hazzan}\\[2pt] \noindent${}$\hfill
\vspace{5mm}

\section{Introduction, Results and Synopsis}
\label{s:IRS}
Supersymmetry has been studies for over forty years\cite{r1001,rWB}, has had successful application in nuclear physics\cite{rSuSyNP80,rSuSyNP10}, critical phenomena\cite{rFQS-SuSy,rGJRSV-SuSyCM}, and has recently found applications also in condensed matter physics: see the recent reviews\cite{rSuSyCM11-PG,rSuSyCM13} for example. In quantum applications, the supermultiplets must be off-shell, \ie, free of any (space)time-differential constraint that could play the role of the Euler-Lagrange (classical) equation of motion. The long-standing challenge of a systematic classification of off-shell supermultiplets\cite{rGLPR,rGLP} has been addressed with significant success in the last decade or so; see Refs.\cite{rBKMO,rKRT,rKT07,rILS,rUMD09-1,rDI-SQM11,rUMD09-2,rGHHS-CLS,rUMD12-3} and references therein. One of the pivotal ideas enabling this recent development was the use of graph-theoretical methods\cite{rA,r6-1,r6-1.2} in assessing the structure of the supersymmetry transformations within off-shell supermultiplets, and which turned out to relate the classification problem to encryption and coding theory\cite{r6-3,r6-3.2,r6-3.1}.

Although this research program uncovered trillions of off-shell supermultiplets of worldline $N$-extended supersymmetry, concurrent research\cite{rDHIL13} shows that this is merely a discrete subset of a vast continuum---which may well come as a surprise, since both the continuous Lie algebras and the various discrete symmetry groups familiar from physics applications all have discrete sequences of unitary, linear and finite-dimensional representations.
 Ref.\cite{rTHGK12} showed that the infinite sequence of quotient supermultiplets specified in Ref.\cite{r6-1} in fact defines an infinite sequence of ever larger unitary, linear and finite-dimensional off-shell representations of $N\,{\geqslant}\,3$-extended worldline supersymmetry, and Ref.\cite{rTHGK13} finds highly non-trivial and intricate dynamics for the simplest of these supermultiplets---even with only bilinear Lagrangians.
 
Herein, we continue this line of research an prove that these supermultiplets are indeed merely special cases of a continuum, which can be physically probed and observed.

For simplicity and concreteness, we focus on the worldsheet $N\,{=}\,4$-extended worldline supersymmetry algebra
\begin{equation}
 \{Q_I,Q_J\}=2i\d_{IJ}\,\vdt ,\qquad
 [\vdt,Q_I]=0,
 \label{e:SuSy}
\end{equation}
where $i\vdt$ is the Hamiltonian (in the familiar $\hbar=1=c$ units) and $Q_1,{\cdots}Q_4$ are the supercharges, four real generators of supersymmetry. We also focus on a particular set of supermultiplets, see\eq{e:QC4} below, which were adapted from Ref.\cite{rTHGK13} by replacing one of the component bosons with its $\t$-derivative and renaming the component fields. Most of our present results then equally apply to the $N\,{=}\,3$ supermultiplet of Refs.\cite{rTHGK12,rTHGK13}. We focus on worldline supersymmetry for several reasons:
 ({\it a})~by dimensional reduction, it is an integral part of any supersymmetric theory,
 ({\it b})~it is directly relevant in diverse fields in physics, from candidates for the fundamental description of $M$-theory\cite{rBBS} to the phenomenology of topological insulators and graphene\cite{rLuHerbut-G}, and
 ({\it c})~it shows up in the Hilbert space of any supersymmetric quantum theory.

\section{The Q-Continuum}
\label{s:QC}
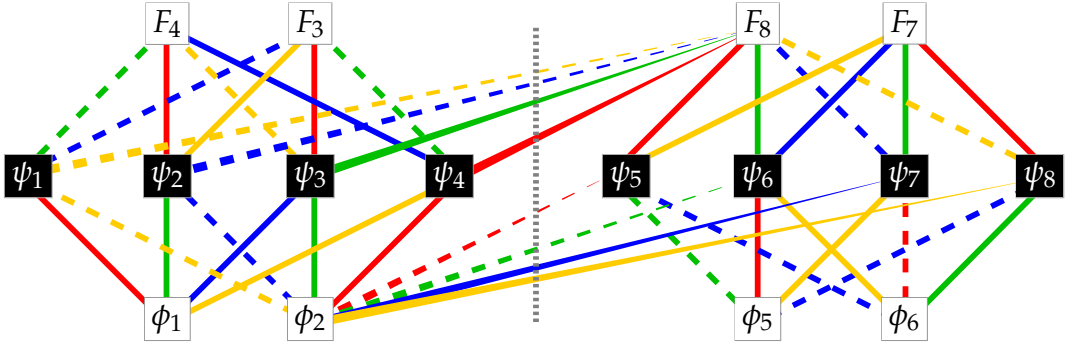
\begin{figure}[ht]
\centering
 \begin{picture}(140,45)
   \put(0,0){\includegraphics[width=140mm]{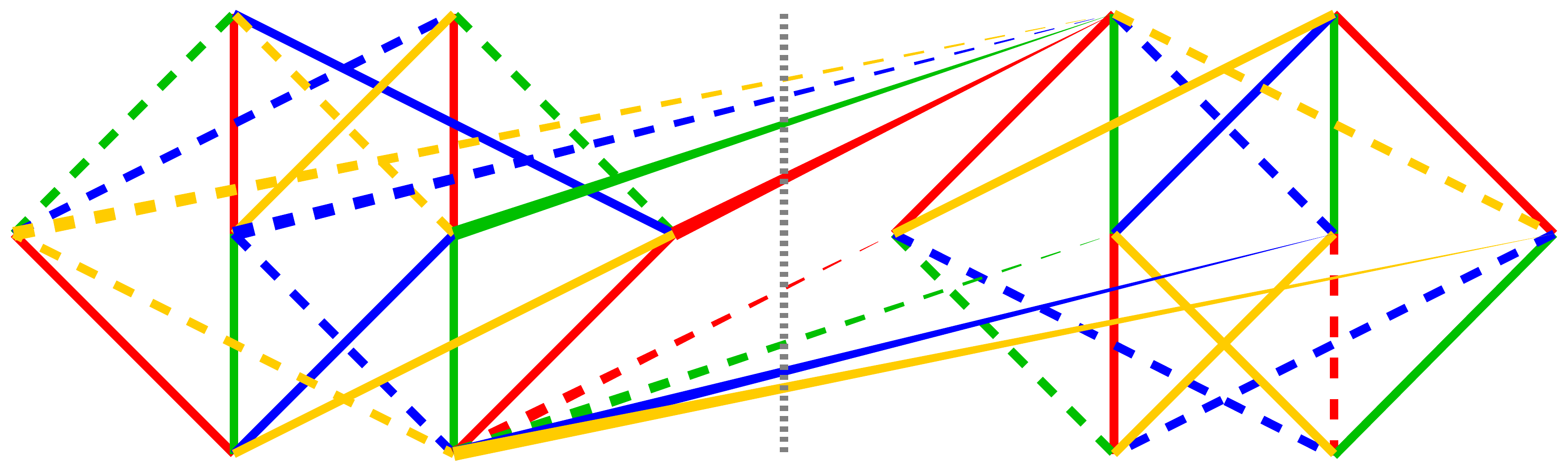}}
    \put(21,1){\cB{$\f_1$}}
    \put(40,1){\cB{$\f_2$}}
    \put(99.5,1){\cB{$\f_5$}}
    \put(119,1){\cB{$\f_6$}}
    \put(2.5,20){\bB{$\j_1$}}
    \put(21,20){\bB{$\j_2$}}
    \put(40,20){\bB{$\j_3$}}
    \put(58.5,20){\bB{$\j_4$}}
    \put(82,20){\bB{$\j_5$}}
    \put(99.5,20){\bB{$\j_6$}}
    \put(119,20){\bB{$\j_7$}}
    \put(137,20){\bB{$\j_8$}}
    \put(21,40){\cB{$F_4$}}
    \put(40,40){\cB{$F_3$}}
    \put(99.5,40){\cB{$F_8$}}
    \put(119,40){\cB{$F_7$}}
 \end{picture}
\caption{A graphical depiction of the $N=4$ worldline supermultiplet\eq{e:QC4}.}
 \label{f:QC4}
\end{figure}

We proceed by way of a concrete example.
 Adapting from Refs.\cite{rTHGK12,rTHGK13}, we study the 1-parameter family of off-shell supermultiplets of worldline $N\,{=}\,4$-extended supersymmetry without central extensions
\begin{equation}
  \begin{array}{@{} c|c@{~~}c@{~~}c@{~~}c @{}}
 & \C3{Q_1} & \C1{Q_2} & \C6{Q_3} & \C2{Q_4} \\ 
    \toprule\rule{0pt}{12pt}
\f_1
 & \j_1 & \j_2 & \j_3 & \j_4 \\[1pt]
\f_2
 & \j_4{-}\C4\a\j_5 & \j_3{-}\C4\a\j_6
  & -\j_2{+}\C4\a\j_7 & -\j_1{+}\C4\a\j_8 \\[1pt]
F_3
 & \dt\j_3 & -\dt\j_4 & -\dt\j_1 & \dt\j_2 \\[1pt]
F_4
 & \dt\j_2 & -\dt\j_1 & \dt\j_4 & -\dt\j_3 \\[3pt]
    \cline{2-5}
\f_5
 & \j_6 & -\j_5   & -\j_8 & \j_7 \\[1pt]
\f_6
 & -\j_7 & \j_8 & -\j_5 & \j_6 \\[1pt]
F_7
 & \dt\j_8 & \dt\j_7 & \dt\j_6 & \dt\j_5 \\[1pt]
F_8
 & \dt\j_5 & \dt\j_6 & -\dt\j_7 & -\dt\j_8 \\
    \bottomrule
  \end{array}
 \quad
  \begin{array}{@{} c|c@{}c@{}c@{}c @{}}
 & \C3{Q_1} & \C1{Q_2} & \C6{Q_3} & \C2{Q_4} \\ 
    \toprule\rule{0pt}{16pt}
\j_1
 & i\dt\f_1 & -iF_4 & -iF_3 & \!-i\dt\f_2{-}i\C4\a F_8\! \\[0pt]
\j_2
 & iF_4 & i\dt\f_1 & \!-i\dt\f_2{-}i\C4\a F_8\! & iF_3 \\[0pt]
\j_3
 & iF_3 & i\dt\f_2{+}i\C4\a F_8 & i\dt\f_1 & -iF_4 \\[0pt]
\j_4
 & i\dt\f_2{+}i\C4\a F_8 & -iF_3 & iF_4  & i\dt\f_1\\[3pt]
    \cline{2-5}\rule{0pt}{2.5ex}
\j_5
 & iF_8 & -i\dt\f_5 & -i\dt\f_6 & iF_7 \\[0pt]
\j_6
 & i\dt\f_5 & iF_8 & iF_7 & i\dt\f_6 \\[0pt]
\j_7
 & -i\dt\f_6 & iF_7 & -iF_8 & i\dt\f_5 \\[0pt]
\j_8
 & iF_7 & i\dt\f_6 & -i\dt\f_5 & -iF_8 \\[0pt]
    \bottomrule
  \end{array}
  \label{e:QC4}
\end{equation}
which may be depicted\ft{Graphical depictions of supersymmetry transformation rules are a time-tested intuitive tool\cite{rFre}, but has been rigorously formalized only recently\cite{r6-1}, and we adopt those conventions.} in the manner of Figure~\ref{f:QC4}: Component fields are depicted as nodes and the $Q$-transformations between them as connecting edges, variously colored to correspond to the four supercharges $Q_I$, and are drawn solid (dashed) to depict the positive (negative) signs in\eq{e:QC4}.
The tapered edges correspond to the one-way $Q$-transformations the ``magnitude'' of which is parametrized by $\C4\a\in\IR$ in the tabulation\eq{e:QC4}. These transformations are ``one-way'' in the sense that, \eg, $\C3{Q_1}(\f_2)$ contains $\j_5$, but $\C3{Q_1}(\j_5)$ does not contain $\f_2$. Nevertheless of course, the supersymmetry algebra relations\eq{e:SuSy} are fully satisfied on every given component field without needing any (space)time differential condition; the supermultiplet\eq{e:QC4} is thus a proper off-shell representation of $N\,{=}\,4$-extended supersymmetry on the worldline.

The special value $\C4\a\,{=}\,0$ corresponds to the decomposition of the supermultiplet\eq{e:QC4} into two $(2|4|2)$-dimensional supermultiplets. In turn, the off-shell supermultiplet\eq{e:QC4} cannot be decomposed as a direct sum of two separate supermultiplets for any $\C4\a\,{\neq}\,0$. We will refer to $\C4\a$ as the ``tuning parameter'' of the 1-parameter family of supermultiplets\eq{e:QC4}.
 In fact, the off-shell supermultiplets of $N\,{=}\,3$-extended supersymmetry considered in Refs.\cite{rTHGK12,rTHGK13} can be similarly generalized to depend on a precisely analogous tuning parameter, dialing the ``magnitude'' of the one-way $Q_I$-transformations connecting the two halves of the supermultiplet; see Figure~\ref{f:QC4}. Those supermultiplets are closely related to the $\C4\a=1$ version of\eq{e:QC4}: except for some renaming of component fields, one merely needs to drop the fourth supersymmetry and replace $F_4\mapsto\f_4=\int\rd F_4$, effectively lowering the corresponding node (top, left) to the bottom level in the graph in Figure~\ref{f:QC4}.

Explicit attempts verify that no component field redefinition can remove the parameter $\C4\a$ from the supersymmetry all of the transformation rules\eq{e:QC4}, so that this table indeed defines a 1-parameter continuum of distinct off-shell representations of worldline $N\,{=}\,4$-extended supersymmetry; it represents one of the optimal choices, in that $\C4\a$ occurs in fewest terms.

\section{Lagrangians}
 \label{s:L}
We now turn to show that the explicit dependence on the tuning parameter $\C4\a$ does show up in the dynamics. To this end, we construct sufficiently general Lagrangians for direct use in classical applications by means of the ensuing Euler-Lagrange equations, or in quantum models via the partition functional
 $Z[\f_*]\Defl\pmb{\int\! D\texttt{[}}\mkern1mu\f\mkern2mu\pmb{\texttt]}
   \exp\{i\!\int\!\rd\t\,L[\f_*{+}\f,\dt\f_*{+}\dt\f,\dots]\}$, or simply using the Hamiltonian, $H\Defl p{\cdot}\dt{q}-L$, corresponding to the chosen Lagrangian.

\subsection{Kinetic Terms}
Following the procedure employed in Ref.\cite{rTHGK13}, we use the fact that any Lagrangian of the form
\begin{equation}
  L \Defl -\C2{Q_4}\C6{Q_3}\C1{Q_2}\C3{Q_1}\,k(\f,\j,F)
 \label{e:GE}
\end{equation}
is automatically supersymmetric, since its $\d_Q\Defl i\e^IQ_I$-transformation necessarily produces a total $\t$-derivative. This is the direct adaptation of the construction of the so-called ``$D$-terms'' in standard treatments of supersymmetry\cite{r1001,rWB}.

Dimensional analysis dictates that for kinetic-type Lagrangians we need $k(\f,\j,F)$ to be bilinear in the component fields $\f_1,\f_2,\f_5,\f_6$; this will produce terms of the form $\dt\f_a\dt\f_b$, $i\j_\a\dt\j_\b$, $F_AF_B$ and $\dt\f_a F_B$, as appropriate for kinetic terms.
 Table~\ref{t:KEaa} lists the individually supersymmetric Lagrangian summands obtained this way, after dropping total $\t$-derivatives. As shown, the ten bilinear functions $k(\f)=k^{a,b}\f_a\f_b$ result in six linearly independent terms, so we define
\begin{equation}
  L^{\sss\text{KE}}_{\sss\vec{A}} \Defl
  -\C2{Q_4}\C6{Q_3}\C1{Q_2}\C3{Q_1}
    \big(\fc12A_1\f_1^{~2} + \fc12A_2\f_2^{~2} + \fc12A_3\f_5^{~2}
        + A_4\f_1\f_2 + A_5\f_1\f_5 + A_6\f_1\f_6\big),
 \label{e:KE}
\end{equation}
and read off the actual summands from Table~\ref{t:KEaa}, to save space.
\begin{table}[tdp]
 \caption{Manifestly supersymmetric kinetic Lagrangian terms for the $\C4\a$-supermultiplet}
\label{t:KEaa}
\vspace*{-5mm}
$$
\begin{array}{@{} r|l @{}}
\bs{\f_i\mkern1mu\f_j}
 & \bs{-Q^4(\f_i\mkern1mu\f_j)
    \Defl-\C2{Q_4}\C6{Q_3}\C1{Q_2}\C3{Q_1}(\f_i\mkern1mu\f_j)}
\\ \toprule
\fc12\f_1^{~2}
&
+(\dt\f_1)^2
+(\dt\f_2{+}\C4\a F_8)^2
+ F_3^{~2}
+ F_4^{~2}
+i \j_1 \dt\j_1 
+i \j_2 \dt\j_2 
+i \j_3 \dt\j_3 
+i \j_4 \dt\j_4 
\\
\fc12\f_2^{~2}
&
+(\dt\f_1{-}\C4\a F_7)^2
+(\dt\f_2)^2
+(F_3{-}\C4\a\dt\f_5)^2
+(F_4{+}\C4\a\dt\f_6)^2
+i(\j_1{-}\C4\a\j_8)(\dt\j_1{-}\C4\a\dt\j_8)
\\[-1pt] &\qquad
+i(\j_2{-}\C4\a\j_7)(\dt\j_2{-}\C4\a\dt\j_7)
+i(\j_3{-}\C4\a\j_6)(\dt\j_3{-}\C4\a\dt\j_6)
+i(\j_4{-}\C4\a\j_5)(\dt\j_4{-}\C4\a\dt\j_5)
\\
\fc12\f_5^{~2}
&
+(\dt\f_5)^2
+(\dt\f_6)^2
+ F_7^{~2}
+ F_8^{~2}
+i \j_5 \dt\j_5 
+i \j_6 \dt\j_6 
+i \j_7 \dt\j_7 
+i \j_8 \dt\j_8 
\\ \midrule
\f_1 \f_2
&
-2 \C4\a \dt\f_1 F_8
+2 \C4\a \dt\f_2 F_7 
-2 \C4\a \dt\f_5 F_4
-2 \C4\a \dt\f_6 F_3
+2 \C4\a^2 F_8 F_7 
\\[-1pt] &\qquad
-2i \C4\a \j_1 \dt\j_5
-2i \C4\a \j_2 \dt\j_6
+2i \C4\a \j_3 \dt\j_7 
+2i \C4\a \j_4 \dt\j_8
\\
\f_1 \f_5
&
+2 \dt\f_1 \dt\f_5 
-2 \dt\f_2 \dt\f_6 
-2 \C4\a \dt\f_6 F_8
-2 F_3 F_7 
-2 F_4 F_8 
+2i \j_1 \dt\j_6 
-2i \j_2 \dt\j_5 
-2i \j_3 \dt\j_8 
+2i \j_4 \dt\j_7 
\\
\f_1 \f_6
&
+2\dt\f_1 \dt\f_6 
+2 \dt\f_2 \dt\f_5 
+2 \C4\a \dt\f_5 F_8
-2 F_3 F_8 
+2 F_4 F_7 
-2i \j_1 \dt\j_7 
+2i \j_2 \dt\j_8 
-2i \j_3 \dt\j_5 
+2i \j_4 \dt\j_6 
\\
 \bottomrule
\MC2l{\text{Also,~~}\rule{0pt}{2.25ex}
        Q^4(\f_6^{~2})\simeq Q^4(\f_5^{~2}),~~
        Q^4(\f_6^{~2})\simeq Q^4(\f_5^{~2}),~~
        Q^4(\f_2\f_6)\simeq Q^4(\f_1\f_5),~~
        Q^4(\f_5\f_6)\simeq 0.}\\
\end{array}
$$
\end{table}
 For example,
\begin{equation}
 \begin{aligned}
  L^{\sss\text{KE}}_{\sss(1,0,1,0,0,0)}
  &=  (\dt\f_1)^2
     +(\dt\f_2{+}\C4\a F_8)^2
     + F_3^{~2}
     + F_4^{~2} 
     +(\dt\f_5)^2
     +(\dt\f_6)^2
     + F_7^{~2}
     + F_8^{~2} \\
   &\mkern30mu
     +i \j_1 \dt\j_1 
     +i \j_2 \dt\j_2 
     +i \j_3 \dt\j_3 
     +i \j_4 \dt\j_4
     +i \j_5 \dt\j_5 
     +i \j_6 \dt\j_6 
     +i \j_7 \dt\j_7 
     +i \j_8 \dt\j_8 
 \end{aligned}
 \label{e:StdKE}
\end{equation}
defines the ``standard-looking'' kinetic terms for this supermultiplet. Throughout the six supersymmetric bilinear terms in Table~\ref{t:KEaa}, the component field $\f_2$ appears only with a derivative acting on it. Therefore---if the Lagrangian were limited to\eq{e:StdKE}---it would be possible to perform the non-local component field redefinition
\begin{equation}
  \f_2~~\mapsto~~F_2 \Defl (\dt\f_2{+}\C4\a F_8)
   \quad\text{and}\quad
  \f_2 = \int\!\rd\t\,(F_2{-}\C4\a F_8),
 \label{e:F2}
\end{equation}
which would also eliminate the appearance of the continuous parameter $\C4\a$ from the ``standard-looking'' Lagrangian\eq{e:StdKE} and would thus {\em\/seem\/} to render the supermultiplets\eq{e:QC4} with various values of the tuning parameter $\C4\a$ equivalent to each other.

However, Table~\ref{t:KEaa} shows that the tuning parameter $\C4\a$ appears in numerous other places and in other field combinations. Also, $\dt\f_2$ appears by itself already in the second row in Table~\ref{t:KEaa}, so that the component field redefinition\eq{e:F2} has the effect
\begin{alignat}9
 -\C2{Q_4}\C6{Q_3}\C1{Q_2}\C3{Q_1}(\fc12\f_1^{~2})
 &=+(\dt\f_1)^2
+(\dt\f_2{+}\C4\a F_8)^2+\dots&\quad
 &\too{\text{(\ref{e:F2})}}
  +(\dt\f_1)^2 +F_2^{~2}+\dots\\
 -\C2{Q_4}\C6{Q_3}\C1{Q_2}\C3{Q_1}(\fc12\f_2^{~2})
 &=+(\dt\f_1)^2
+(\dt\f_2{+}\C4\a F_8)^2+\dots&\quad
 &\too{\text{(\ref{e:F2})}}
  +(\dt\f_1{-}\C4\a F_7)^2 +(F_2{-}\C4\a F_8)^2+\dots
\end{alignat}
of merely shifting the appearance of the tuning parameter $\C4\a$ from one place to another.
 Indeed, no component field redefinition can remove the dependence on the continuous parameter $\C4\a$ from the {\em\/general\/} Lagrangian---even if restricted to just the bilinear terms in Table~\ref{t:KEaa}.

Many of the summands in the lower portion of Table~\ref{t:KEaa} have negative signs, and so would---if used on their own---contribute negatively to the kinetic energy, \ie, induce non-positivity of the kinetic energy and non-unitarity in general. However, when used with the first three supersymmetric sets of kinetic terms (which are positive-definite), it is clear that unitarity constrains the coefficients $A_i$ in\eq{e:KE} so that $A_4,A_5,A_6$ should be sufficiently smaller than $A_1,A_2,A_3$. This is similar to the analogous case examined in Ref.\cite{rTHGK13}.

The conclusion is that there remains a 6-dimensional open neighborhood of Lagrangians that do define unitary quantum models, and most of such models depend explicitly on the tuning parameter $\C4\a$.
 This parameter $\C4\a$ then must be a genuine, observable characteristic of the supermultiplet\eq{e:QC4}. We conclude that the supermultiplets\eq{e:QC4} which differ only in a different choice of the parameter $\C4\a$ cannot be regarded as physically equivalent in general.
 This dependence on the tuning parameter $\C4\a$ becomes only more complex in the general ``$D$-term'' Lagrangians\eq{e:GE}.

In fact, we can strengthen this result as follows.
 Consider two separate supermultiplets of the type\eq{e:QC4}, and label their separate continuous tuning parameters $\C4\a$ and $\C4\b$, respectively:
\begin{equation}
  (\f_1,\f_2,\f_5,\f_6|\j_1,{\cdots}\j_8|F_3,F_4,F_7,F_8)_{\C4\a}
   \quad\text{and}\quad
  (\vf_1,\vf_2,\vf_5,\vf_6|\c_1,{\cdots}\c_8|G_3,G_4,G_7,G_8)_{\C4\b}.
 \label{e:ab}
\end{equation}
Now consider even just the bilinear coupling Lagrangians of the form:
\begin{equation}
  L^{\sss\text{KE}}_{\sss\vec{A};\,\C4\a,\C4\b} \Defl
  -\C2{Q_4}\C6{Q_3}\C1{Q_2}\C3{Q_1}
    \big( \vec{A}^{\sss(\C4\a)}{\cdot}\vec{f}(\f)
         + \vec{A}^{\sss(\C4\b)}{\cdot}\vec{g}(\vf)
          + \vec{A}^{\sss(\C4\a,\C4\b)}{\cdot}\vec{h}(\f,\vf) \big),
 \label{e:KEab}
\end{equation}
where $\vec{A}^{\sss(\C4\a)}{\cdot}\vec{f}(\f)$ are the bilinear terms\eq{e:KE} and the $\vec{A}^{\sss(\C4\b)}{\cdot}\vec{g}(\vf)$ terms are constructed in a precisely analogously way but for the supermultiplet $(\vf|\c|G)_{\C4\b}$, of course with an independent set of six coefficients. Finally, $\vec{A}^{\sss(\C4\a,\C4\b)}{\cdot}\vec{h}(\f,\vf)$ represents the mixing terms, constructed as a general linear combination of the fourteen analogously constructed terms, listed in Table~\ref{t:KEab}.
\begin{table}[!htdb]
 \caption{Fourteen bilinear ``$D$-term''-type manifestly supersymmetric Lagrangian terms that couple the $\C4\a$-supermultiplet with the $\C4\b$-supermultiplet}
\label{t:KEab}
\vspace*{-5mm}
$$
\begin{array}{@{} r|l @{}}
\bs{\f_i\mkern2mu\vf_j}
 &\bs{-Q^4(\f_i\mkern2mu\vf_j) \Defl
       -\C2{Q_4}\C6{Q_3}\C1{Q_2}\C3{Q_1}(\f_i\mkern2mu\vf_j)}
\\ \toprule
\f_1 \vf_1
&
+2 \dt\f_1 \dt\vf_1 
+2(\dt\f_2{+}\C4\a F_8)(\dt\vf_2{+}\C4\b G_8)
+2 F_3 G_3 
+2 F_4 G_4 
~+~
 2i \j_1 \dt\c_1 
+2i \j_2 \dt\c_2 
+2i \j_3 \dt\c_3 
+2i \j_4 \dt\c_4 
\\[1mm]
\f_1 \vf_2
&
+2 \dt\f_1 \dt\vf_2 
-2(\dt\f_2{+}\C4\a F_8)(\dt\vf_1{-}\C4\b G_7)
-2 F_3(G_4{+}\C4\b\dt\vf_6)
+2 F_4(G_3{-}\C4\b\dt\vf_5)
\\
&~~
+2i \j_1(\dt\c_4{-}\C4\b\dt\c_5)
+2i \j_2(\dt\c_3{-}\C4\b\dt\c_6)
-2i \j_3(\dt\c_2{-}\C4\b\dt\c_7)
-2i \j_4(\dt\c_1{-}\C4\b\dt\c_8)
\\[1mm]
\f_1 \vf_5
&
+2 \dt\f_1 \dt\vf_5 
-2(\dt\f_2{+}\C4\a F_8)\dt\vf_6 
-2 F_3 G_7 
-2 F_4 G_8 
~+~
 2i \j_1 \dt\c_6 
-2i \j_2 \dt\c_5 
-2i \j_3 \dt\c_8 
+2i \j_4 \dt\c_7 
\\[1mm]
\f_1 \vf_6
&
+2 \dt\f_1 \dt\vf_6 
+2(\dt\f_2{+}\C4\a F_8)\dt\vf_5 
-2 F_3 G_8 
+2 F_4 G_7 
~-~
 2i \j_1 \dt\c_7 
+2i \j_2 \dt\c_8 
-2i \j_3 \dt\c_5 
+2i \j_4 \dt\c_6 
\\[1mm]
\f_2 \vf_1
&
+2 \dt\f_2 \dt\vf_1 
-2(\dt\f_1{-}\C4\a F_7)(\dt\vf_2{+}\C4\b G_8)
+2(F_3{-}\C4\a\dt\f_5)G_4 
-2(F_4{+}\C4\a\dt\f_6)G_3 
\\
&~~
+2i(\j_4{-}\C4\a\j_5)\dt\c_1 
+2i(\j_3{-}\C4\a\j_6)\dt\c_2 
-2i(\j_2{-}\C4\a\j_7)\dt\c_3 
-2i(\j_1{-}\C4\a\j_8)\dt\c_4 
\\[1mm]
\f_2 \vf_2
&
+2(\dt\f_1{-}\C4\a F_7)(\dt\vf_1{-}\C4\b G_7)
+2 \dt\f_2 \dt\vf_2 
+2(F_3{-}\C4\a\dt\f_5)(G_3{-}\C4\b\dt\vf_5)
+2(F_4{+}\C4\a\dt\f_6)(G_4{+}\C4\b\dt\vf_6)
\\
&~~
+2i(\j_1{-}\C4\a\j_8)(\dt\c_1{-}\C4\b\dt\c_8)
+2i(\j_2{-}\C4\a\j_7)(\dt\c_2{-}\C4\b\dt\c_7)
\\
&~~
+2i(\j_3{-}\C4\a\j_6)(\dt\c_3{-}\C4\b\dt\c_6)
+2i(\j_4{-}\C4\a\j_5)(\dt\c_4{-}\C4\b\dt\c_5)
\\[1mm]
\f_2 \vf_5
&
+2 \dt\f_2 \dt\vf_5 
+2(\dt\f_1{-}\C4\a F_7)\dt\vf_6 
-2(F_3{-}\C4\a\dt\f_5)G_8 
+2(F_4{+}\C4\a\dt\f_6)G_7 
\\
&~~
-2i(\j_1{-}\C4\a\j_8)\dt\c_7 
+2i(\j_2{-}\C4\a\j_7)\dt\c_8 
-2i(\j_3{-}\C4\a\j_6)\dt\c_5 
+2i(\j_4{-}\C4\a\j_5)\dt\c_6 
\\[1mm]
\f_2 \vf_6
&
+2 \dt\f_2 \dt\vf_6 
-2(\dt\f_1{-}\C4\a F_7)\dt\vf_5 
+2(F_3{-}\C4\a\dt\f_5)G_7 
+2(F_4{+}\C4\a\dt\f_6)G_8 
\\
&~~
-2i(\j_1{-}\C4\a\j_8)\dt\c_6 
+2i(\j_2{-}\C4\a\j_7)\dt\c_5 
+2i(\j_3{-}\C4\a\j_6)\dt\c_8 
-2i(\j_4{-}\C4\a\j_5)\dt\c_7 
\\[1mm]
\f_5 \vf_1
&
+2 \dt\f_5 \dt\vf_1 
-2 \dt\f_6(\dt\vf_2{+}\C4\b G_8)
-2 F_7 G_3 
-2 F_8 G_4 
~-~
 2i \j_5 \dt\c_2 
+2i \j_6 \dt\c_1 
+2i \j_7 \dt\c_4 
-2i \j_8 \dt\c_3 
\\[1mm]
\f_5 \vf_2
&
+2 \dt\f_5 \dt\vf_2 
+2 \dt\f_6(\dt\vf_1{-}\C4\b G_7)
+2 F_7(G_4{+}\C4\b\dt\vf_6)
-2 F_8(G_3{-}\C4\b\dt\vf_5)
\\
&~~
-2i \j_5(\dt\c_3{-}\C4\b\dt\c_6)
+2i \j_6(\dt\c_4{-}\C4\b\dt\c_5)
-2i \j_7(\dt\c_1{-}\C4\b\dt\c_8)
+2i \j_8(\dt\c_2{-}\C4\b\dt\c_7)
\\[1mm]
\f_5 \vf_5
&
+2 \dt\f_5 \dt\vf_5 
+2 \dt\f_6 \dt\vf_6 
+2 F_7 G_7 
+2 F_8 G_8 
~+~
 2i \j_5 \dt\c_5 
+2i \j_6 \dt\c_6 
+2i \j_7 \dt\c_7 
+2i \j_8 \dt\c_8 
\\[1mm]
\f_5 \vf_6
&
+2 \dt\f_5 \dt\vf_6 
-2 \dt\f_6 \dt\vf_5 
+2 F_7 G_8 
-2 F_8 G_7 
~-~
 2i \j_5 \dt\c_8 
-2i \j_6 \dt\c_7 
+2i \j_7 \dt\c_6 
+2i \j_8 \dt\c_5 
\\[1mm]
\f_6 \vf_1
&
+2 \dt\f_6 \dt\vf_1 
+2 \dt\f_5(\dt\vf_2{+}\C4\b G_8)
+2 F_7 G_4 
-2 F_8 G_3 
~-~
 2i \j_5 \dt\c_3 
+2i \j_6 \dt\c_4 
-2i \j_7 \dt\c_1 
+2i \j_8 \dt\c_2 
\\[1mm]
\f_6 \vf_2
&
+2 \dt\f_6 \dt\vf_2 
-2 \dt\f_5(\dt\vf_1{-}\C4\b G_7)
+2 F_7(G_3{-}\C4\b\dt\vf_5)
+2 F_8(G_4{+}\C4\b\dt\vf_6)
\\
&~~
+2i \j_5(\dt\c_2{-}\C4\b\dt\c_7)
-2i \j_6(\dt\c_1{-}\C4\b\dt\c_8)
-2i \j_7(\dt\c_4{-}\C4\b\dt\c_5)
+2i \j_8(\dt\c_3{-}\C4\b\dt\c_6)
\\ \bottomrule
\MC2l{\text{Also,~~}\rule{0pt}{2.25ex}
  Q^4(\f_6\mkern2mu\vf_5) \simeq Q^4(\f_5\mkern2mu\vf_6),~~
  Q^4(\f_6\mkern2mu\vf_6) \simeq Q^4(\f_5\mkern2mu\vf_5).}
\end{array}
$$
\end{table}

The expression\eq{e:KEab} then provides a $6{+}6{+}14\,{=}\,26$-parameter continuous family of bilinear Lagrangians for the two distinct 1-parameter families of supermultiplets. Generic choices in the 26-dimensional parameter space $\{\vec{A}^{\sss(\C4\a)},\vec{A}^{\sss(\C4\b)},\vec{A}^{\sss(\C4\a,\C4\b)}\}$ define Lagrangians that depend irremovably on both the tuning parameters $\C4\a$ and $\C4\b$, and so provide for dynamical responses that can be used to observe the values of $\C4\a$ and $\C4\b$, and indeed any difference between them. This then is the practical distinction between $(\f|\j|F)_{\C4\a}$ and $(\vf|\c|G)_{\C4\b}$, which makes these two off-shell representations of $N\,{=}\,4$-extended supersymmetry---as well as any other member of the continuum of supermultiplets\eq{e:QC4}---all {\em\/usefully inequivalent\/} in the sense of Ref.\cite{rChiLin}.

Since $\C4\a$ and $\C4\b$ may be continuously varied, the existence of the coupling Lagrangian\eq{e:KEab}, even if merely bilinear, proves that the members of the continuum of worldline off-shell supermultiplets\eq{e:QC4} are usefully inequivalent. Incidentally, the same can be shown for the $\C4\a\neq1$ versions of the $N\,{=}\,3$ supermultiplets studied in Refs.\cite{rTHGK12,rTHGK13}.

\subsection{Super-Zeeman Terms}
We now turn to Lagrangian terms that are still bilinear, but where dimensional analysis requires an overall dimension-full parameter, of the kind that may be identified as a Larmor-like frequency, coupling the supermultiplet\eq{e:QC4} to external magnetic fields\cite{r6-7a,rTHGK13}.

In general, we seek functions $f(\f,\j,F)$ such that each of
\begin{equation}
  \C6{Q_3}\C1{Q_2}\C3{Q_1}\,f(\f,\j,F),\quad
  \C2{Q_4}\C1{Q_2}\C3{Q_1}\,f(\f,\j,F),\quad
  \C2{Q_4}\C6{Q_3}\C3{Q_1}\,f(\f,\j,F),\quad
  \C2{Q_4}\C6{Q_3}\C1{Q_2}\,f(\f,\j,F)
 \label{e:4Zs}
\end{equation}
vanishes modulo total derivatives. Then, the six quadratic derivatives
\begin{equation}
 \begin{gathered}
  \C1{Q_2}\C3{Q_1}\,f(\f,\j,F),\quad
  \C6{Q_3}\C3{Q_1}\,f(\f,\j,F),\quad
  \C6{Q_3}\C1{Q_2}\,f(\f,\j,F),\\
  \C2{Q_4}\C3{Q_1}\,f(\f,\j,F),\quad
  \C2{Q_4}\C1{Q_2}\,f(\f,\j,F),\quad
  \C2{Q_4}\C6{Q_3}\,f(\f,\j,F)
\end{gathered}
 \label{e:6Zs}
\end{equation}
are all manifestly supersymmetric: When applying $\d_Q=i\e{\cdot}Q$, the $Q_I$ from $\d_Q$ either equals one of the two $Q_I$'s used in the definition\eq{e:6Zs} and so produces $i\vdt$ by\eq{e:SuSy}, or it doesn't and so reproduces one of the expressions\eq{e:4Zs} and again a total $\t$-derivative by assumption\eq{e:4Zs}.
\begin{table}[htdp]
 \caption{The $\C6{Q_3}\C1{Q_2}\C3{Q_1}$-transforms of bosonic bilinear terms, modulo total $\t$-derivatives}
\label{t:QZaa}
\vspace*{-5mm}
$$
\begin{array}{@{} r|l @{}}
\bs{\f_i\mkern1mu\f_j} & -i\bs{\C6{Q_3}\C1{Q_2}\C3{Q_1}(\f_i\mkern1mu\f_j)}
\\ \toprule
\fc12\f_1 \f_1
&
 + \dt\f_1 \j_4
 -(\dt\f_2{+}\C4\a F_8)\j_1
 + F_3 \j_2
 - F_4 \j_3
\\ 
 \fc12\f_2 \f_2
&
 +(\dt\f_1{-}\C4\a F_7)(\j_4{-}\C4\a\j_5)
 - \dt\f_2(\j_1{-}\C4\a\j_8)
 +(F_3{-}\C4\a\dt\f_5)(\j_2{-}\C4\a\j_7)
 -(F_4{+}\C4\a\dt\f_6)(\j_3{-}\C4\a\j_6)
\\ 
 \fc12\f_5 \f_5
&
 + \dt\f_5 \j_7
 + \dt\f_6 \j_6
 + F_7 \j_5
 - F_8 \j_8
\\ 
 \fc12\f_6 \f_6
&
 + \dt\f_5 \j_7
 + \dt\f_6 \j_6
 + F_7 \j_5
 - F_8 \j_8 \smash{\raisebox{1.2ex}{$\left.\rule{0pt}{2.4ex}\right\}$~subtract}}
\\[1mm]
 \f_1 \f_2
&
 +\C4\a[\dt\f_1 \j_8
 +(\dt\f_2{+}\C4\a F_8)\j_5
 - F_3 \j_6
 - F_4 \j_7
 + \dt\f_5 \j_3
 - \dt\f_6 \j_2
 - F_7 \j_1
 - F_8 \j_4]
\\ 
 \f_1 \f_5
&
 + \dt\f_1 \j_7
 -(\dt\f_2{+}\C4\a F_8)\j_6
 - F_3 \j_5
 + F_4 \j_8
 + \dt\f_5 \j_4
 + \dt\f_6 \j_1
 - F_7 \j_2
 + F_8 \j_3
\\ 
 \f_2 \f_6
&
 - \dt\f_1 \j_7
 +(\dt\f_2{+}\C4\a F_8)\j_6
 + F_3 \j_5
 - F_4 \j_8
 - \dt\f_5 \j_4
 - \dt\f_6 \j_1
 + F_7 \j_2
 - F_8 \j_3 \smash{\raisebox{1.2ex}{$\left.\rule{0pt}{2.4ex}\right\}$~add}}
\\ 
 \f_1 \f_6
&
 + \dt\f_1 \j_6
 +(\dt\f_2{+}\C4\a F_8)\j_7
 + F_3 \j_8
 + F_4 \j_5
 - \dt\f_5 \j_1
 + \dt\f_6 \j_4
 - F_7 \j_3
 - F_8 \j_2
\\ 
 \f_2 \f_5
&
 + \dt\f_1 \j_6
 +(\dt\f_2{+}\C4\a F_8)\j_7
 + F_3 \j_8
 + F_4 \j_5
 - \dt\f_5 \j_1
 + \dt\f_6 \j_4
 - F_7 \j_3
 - F_8 \j_2 \smash{\raisebox{1.2ex}{$\left.\rule{0pt}{2.4ex}\right\}$~subtract}}
\\ 
 \f_5 \f_6
&-\vdt(\f_5\j_6{+}\f_6\j_7)\simeq0 
\\ \bottomrule\end{array}
$$
\end{table}
Such terms remind of the so-called ``$F$-terms'' in standard treatments of supersymmetry\cite{r1001,rWB}.

We again restrict to bilinear terms for simplicity, and Table~\ref{t:QZaa} presents the linearly independent such terms, obtained applying only the first batch of three supercharges. The other three expressions\eq{e:4Zs} each produce analogous results with a pattern virtually identical to the one shown in Table~\ref{t:QZaa}. The last-row entry, $\f_5\f_6$, results in a total $\t$-derivative all by itself, and simple row-operations (indicated by braces) show that we can form three more. This means that each of the twenty-four terms
\begin{equation}
  \frc12 Q_IQ_J(\f_5^{~2}-\f_6^{~2}),\quad
  Q_IQ_J(\f_1\f_5+\f_2\f_6),\quad
  Q_IQ_J(\f_1\f_6-\f_2\f_5),\quad
  Q_IQ_J(\f_5\f_6)
\end{equation}
is a supersymmetric Lagrangian contribution. This list turns out repetitive, and contains only four linearly independent expressions, listed in Table~\ref{t:Zaa}.
\begin{table}[htdp]
 \caption{Super-Zeeman bilinear contributions, modulo total $\t$-derivatives}
\label{t:Zaa}
\vspace*{-6mm}
$$
\begin{array}{@{} r@{\,\Defl\,}l @{}}
\toprule
Z_1& \f_5 F_7 -\f_6 F_8 +i\j_5\j_7 -i\j_5\j_7  
\\
Z_2& \f_5 F_8 +\f_6 F_7 +i\j_5\j_6 +i\j_7\j_8
\\
Z_3& \f_1 F_7 +\f_2 F_8 -\f_5 F_3 +\f_6 F_4 
 -i\j_1\j_8 -i\j_2\j_7 -i\j_3\j_6 -i\j_4\j_5
\\
Z_4& \f_1 F_8 -\f_2 F_7 -\f_6 F_3 -\f_5 F_4
 -i\j_1\j_5 -i\j_2\j_6 +i\j_3\j_7 +i\j_4\j_8
 -\C4\a(\f_5 \dt\f_6 +i\j_5\j_8 +i\j_6\j_7)
\\
\bottomrule
\end{array}
$$
\end{table}
The most general super-Zeeman type Lagrangian bilinear in the component fields of the $(\f|\j|F)_{\C4\a}$ supermultiplet is therefore
\begin{equation}
  L^{\sss\text{SZ}}_{\sss\vec{B};\,\C4\a} \Defl
   B_1 Z_1 + B_2 Z_2 + B_3 Z_3 + B_4 Z_4,
 \label{e:BZ}
\end{equation}
with the terms $Z_i$ listed in Table~\ref{t:Zaa}.
Of these, only the last term contains the expression
\begin{equation}
  B_5Z_5 =\dots -\C4\a B_5\f_5\dt\f_6+\dots
      \simeq\dots -\frc12\C4\a B_5(\f_5\dt\f_6-\dt\f_5\f_6)+\dots
 \label{e:f56}
\end{equation}
which in Lagrangian physics may be interpreted as the coupling of the magnetic field $B_5$ to the angular momentum of rotation in the $(\f_5,\f_6)$-plane---if the bosons $\f_5,\f_6$ are interpreted as Cartesian coordinates in the target space. 
 The elimination of the auxiliary fields $F_3,F_4,F_7,F_8$ (and $G_3,G_4,G_7,G_8$) by means of their equations of motion however induces many additional terms of the type\eq{e:f56}. This justifies the identification of the terms\eq{e:BZ} and the supersymmetric version of the $\vec{B}{\cdot}\vec{L}$ terms exhibiting the Zeeman effect.

The four terms in Table~\ref{t:Zaa} together with their $(\f|\j|F)_{\C4\a}\to(\vf|\c|G)_{\C4\b}$ counterparts and the 26-parameter Lagrangian\eq{e:KEab} then form the most general, 34-parameter family of bilinear Lagrangians
\begin{equation}
  L^{\sss\text{KE}}_{\sss\vec{A};\,\C4\a,\C4\b}
  +L^{\sss\text{SZ}}_{\sss\vec{B};\,\C4\a}
   +L^{\sss\text{SZ}}_{\sss\vec{B};\,\C4\b}
 \label{e:KESZ}
\end{equation}
for two different supermultiplets from the family\eq{e:QC4}.
 Insuring positivity of the kinetic energy, and unitarity more generally, restricts these parameters to an open neighborhood in this 34-dimensional parameter space.

\section{Sample Dynamics}
 \label{s:Mix}
To illustrate the dependence on $\C4\a$, consider the Lagrangian\eq{e:KE}, with $\vec{A}'=(\fc12a_1,\fc12a_2,\fc12a_3,0,0,0)$, and focus only on the bosonic fields:
\begin{equation}
\begin{aligned}
  L^{\sss\text{KE}}_{\sss\vec{A}';\,\C4\a}
  &=\frc{a_1}2\big[\dt\f_1^{~2}+(\dt\f_2+\C4\a F_8)^2+F_3^{~2}+F_4^{~2}\big]
  \\ &\mkern20mu
    {+}\frc{a_2}2\big[(\dt\f_1{-}\C4\a F_7)^2+\dt\f_2^{~2}
                 +(F_3{-}\C4\a\dt\f_5)^2+(F_4{+}\C4\a\dt\f_6)^2\big]
     +\frc{a_3}2\big[\dt\f_5^{~2}+\dt\f_6^{~2}+F_8^{~2}+F_7^{~2}\big] +\dots
 \end{aligned}
 \label{e:Labc}
\end{equation}
where the ellipses denote the omitted fermionic terms.
 The Euler-Lagrange equations of motion for $F_3,F_4,F_7,F_8$ are of course algebraic and produce the classical relationships such as $F_8=\big(\frac{\C4\a\,a_1}{\C4{\a^2}\,a_1+a_3}\big)\dt\f_2$. Substituting these back into the Lagrangian yields
\begin{equation}
  L^{\sss\text{KE}}_{\sss\vec{A}';\,\C4\a}\big|{}^{}_{\sss F_A}
  =\frc12\big(a_1{+}\frc{a_2a_3}{\C4{\a^2}a_2+a_3}\big)\dt\f_1^{~2}
   +\frc12\big(a_2{+}\frc{a_1a_3}{\C4{\a^2}a_1+a_3}\big)\dt\f_2^{~2}
    +\frc12\big(a_3{+}\frc{\C4{\a^2}a_1a_2}{a_1+a_2}\big)\dt\f_5^{~2}
     +\frc12\big(a_3{+}\frc{\C4{\a^2}a_1a_2}{a_1+a_2}\big)\dt\f_6^{~2}
      +\dots
\end{equation}
changing the effective masses of the bosonic fields $\f_1,\f_2,\f_5,\f_6$---and in three different ways, breaking the usual supersymmetric degeneracy\ft{After the auxiliary fields $F_3,F_4,F_7,F_8$ have been eliminated, even the number of bosons (now four) and fermions (still eight) is unequal. A detailed study of the normal modes is necessary to ascertain if supersymmetry is broken or not, and that can typically only be done for concrete choices of most of the numerous parameters involved; see for example Ref.\cite{r6-7a}. We defer this to a later effort.}. In turn, the 3-parameter family of Hamiltonians computed from the Lagrangians\eq{e:Labc} turns out to be:
\begin{align}
   H_{\sss\vec{A}';\,\C4\a}
   &= \frc{\p_1^{~2}}{2(a_1{+}a_2)} + \frc{\C4\a\,a_2\,F_7}{a_1{+}a_2}\p_1
      -\frc12\big(a_3{+}\frc{\C4{\a^2}a_1a_2}{a_1{+}a_2}\big)F_7^{~2}
     +\frc{\p_2^{~2}}{2(a_1{+}a_2)} - \frc{\C4\a\,a_1\,F_8}{a_1{+}a_2}\p_2
      -\frc12\big(a_3{+}\frc{\C4{\a^2}a_1a_2}{a_1{+}b_2}\big)F_8^{~2}
 \label{e:Haa}
  \\ &\mkern20mu
     {+}\frc{\p_5^{~2}}{2(\C4{\a^2}a_2{+}a_3)}
     + \frc{\C4\a\,a_2\,F_3}{\C4{\a^2}a_2{+}a_3}\p_5
      -\frc12\big(a_1{+}\frc{a_2a_3}{\C4{\a^2}a_2{+}a_3}\big)F_3^{~2}
     +\frc{\p_6^{~2}}{2(\C4{\a^2}a_2{+}a_3)}
      -\frc{\C4\a\,a_2\,F_4}{\C4{\a^2}a_2{+}a_3}\p_6
      -\frc12\big(a_1{+}\frc{a_2a_3}{\C4{\a^2}a_2{+}a_3}\big)F_4^{~2},\nn
\end{align}
where
\begin{equation}
 \begin{aligned}
   \p_1&\Defl(a_1{+}a_2)\dt\f_1{-}\C4\a bF_7,&
   \p_2&\Defl(a_1{+}a_2)\dt\f_2{+}\C4\a aF_8,\\
   \p_5&\Defl(\C4{\a^2}a_2{+}a_3)\dt\f_5{-}\C4\a bF_3,&
   \p_6&\Defl(\C4{\a^2}a_2{+}a_3)\dt\f_6{+}\C4\a aF_4
 \end{aligned}
\end{equation}
are the canonically conjugate momenta.
 The Hamiltonian\eq{e:Haa} may therefore be understood as the auxiliary fields $F_A$ parametrizing a viscous damping/boosting, depending on the sign of the particular $F_A\p_a$-term, the magnitude/intensity of which is controlled by the tuning parameter $\C4\a$.

In turn, the Lagrangian
\begin{equation}
  L^{\sss\text{KE}}_{\sss\vec{A}';\,\C4\a} + B_5 Z_5
\end{equation}
describes a typical coupling to an external magnetic field $B_5$. For example, the corresponding equations of motion for $(\f_5,\f_6)$ are
\begin{equation}
  a_3\ddt\f_5 +\C4\a B_5 \dt\f_6 = 0 = a_3\ddt\f_6 -\C4\a B_5 \dt\f_5,
\end{equation}
which have standard oscillatory solutions
\begin{equation}
  \f_5 = C_1 +C_2\cos(\w\t) +C_3\sin(\w\t)\quad\text{and}\quad
  \f_6 = C_4 -C_3\cos(\w\t) +C_2\sin(\w\t),
\end{equation}
where $\w\Defl\frac{\C4\a B_5}{a_3}$ is the standard Larmor-like frequency of magnetically induced rotations in the $(\f_5,\f_6)$-plane. Quantum-mechanically, the energy levels in this system will become split by integral multiples of $\hbar\w=\hbar\frc{\C4\a B_5}{a_3}$, which is proportional to both the magnitude of the external magnetic field as usual in the Zeeman effect, and also to the tuning parameter $\C4\a$. Distinct supermultiplets\eq{e:ab} then respond to this external magnetic field differently, and a system with two such distinct supermultiplets will respond to the external magnetic field in a way that detects the difference $(\C4\a\neq\C4\b)$.
 This proves that the value of the tuning parameter $\C4\a$ is physically observable, and that distinctly ``tuned'' supermultiplets of the type\eq{e:QC4} are usefully inequivalent\cite{rChiLin}.

\section{Conclusions}
\label{s:Coda}
We have presented a 1-parameter continuous family of off-shell supermultiplets\eq{e:QC4} of $N\,{=}\,4$ worldline supersymmetry, which is closely related to the off-shell supermultiplets studied in Refs.\cite{rTHGK12,rTHGK13}. In fact, all of the qualitative conclusions from study of\eq{e:QC4} apply just as well to those $N\,{=}\,3$ off-shell supermultiplets.

The supermultiplet\eq{e:QC4} exhibits an explicit, continuously variable tuning parameter, labeled $\C4\a$, the value of which controls the relative ``magnitude'' in the binomial results of applying the supercharges to the component fields. By virtue of the existence of these binomial terms, the supermultiplet\eq{e:QC4} may be thought of as a network of Adinkras\cite{r6-1} connected by one-way edges, as depicted in Figure~\ref{f:QC4}.

For two distinct members from this continuous family of off-shell supermultiplets, we have constructed a 34-parameter family of general bilinear Lagrangians\eq{e:KESZ} which:
\begin{enumerate}\itemsep=-3pt\vspace{-2mm}
 \item generalize the ``standard'' kinetic terms\eq{e:StdKE} into a 6-parameter family of Lagrangians\eq{e:KE},
 \item mix two off-shell supermultiplets of the same type\eq{e:QC4}, each with a different value of the tuning parameter, given as 14-parameter linear combinations of the terms from Table~\ref{t:KEab}, and
 \item couple such supermultiplets to external magnetic fields inducing a variant of the super-Zeeman effect, given as 4-parameter linear combinations of the terms from Table~\ref{t:Zaa}.
\end{enumerate}\vspace*{-3mm}
Using the constructions described in Section~\ref{s:L}, these Lagrangians can be generalized to include:
 ({\it a})~higher order interaction terms, and
 ({\it b})~couplings to additional and all differently tuned supermultiplets from the family\eq{e:QC4}.

Section~\ref{s:Mix} then demonstrates that the multi-dimensional parameter space of these Lagrangians admits an open neighborhood where the kinetic energy is guaranteed to be positive, indicating unitarity in the corresponding quantum theory. Except for very special choices within this parameter space, the Lagrangians explicitly depend on the tuning parameter $\C4\a$, and also $\C4\b$ in\eq{e:KEab}, of the supermultiplet\eq{e:QC4}, and in ways that have direct dynamical consequences, and observably affect the response of these supermultiplets to probing by external magnetic fields.

Furthermore, the wealth and diversity of even just the bilinear coupling/mixing terms listed in Table~\ref{t:KEab} indicates that supermultiplets with a different choice of the tuning parameter are indeed observably different, and so usefully inequivalent in the sense of Ref.\cite{rChiLin}. As the same analysis applies just as well for the infinite sequence of supermultiplets discussed in Ref.\cite{rTHGK12}, we thus have clear proof that off-shell supermultiplets of worldline $N$-extended supersymmetry without central extensions form a physically observable continuum.

\bigskip
\bigskip
\paragraph{\bfseries Acknowledgments:}
 TH is grateful to
 the Department of Physics, University of Central Florida, Orlando FL,
 and
 the Physics Department of the Faculty of Natural Sciences of the University of Novi Sad, Serbia, 
 for the recurring hospitality and resources.
 GK is grateful to
 the Department of Physics and Astronomy, Howard University, Washington DC, 
 for the hospitality and resources.

\bigskip
\providecommand{\href}[2]{#2}\begingroup\raggedright
\endgroup
\end{document}